\documentclass[runningheads]{llncs}

\PassOptionsToPackage{numbers, compress}{natbib}

\usepackage[utf8]{inputenc} 
\usepackage[T1]{fontenc}    

\usepackage{url}            
\usepackage{booktabs}       
\usepackage{amsfonts}       
\usepackage{nicefrac}       
\usepackage{microtype}      
\usepackage[sectionbib]{natbib}

\usepackage{multirow}
\usepackage{textcomp}

\usepackage{graphicx}
\usepackage{caption}
\usepackage{subcaption}

\usepackage{float}

\usepackage{algorithm}
\usepackage{algpseudocode}
\usepackage[colorinlistoftodos]{todonotes}

\usepackage{amsmath,amssymb,url,longtable,booktabs,multicol,multirow,lineno}

\usepackage[multiple]{footmisc}
\usepackage{tikz}
\usepackage{tikzscale}
\usepackage{xspace}
\usepackage{pgfplots}

\usepackage{marvosym}

\DeclareUnicodeCharacter{20AC}{\EUR{}}
\usepgfplotslibrary{groupplots}
\pgfplotsset{compat=1.16}
\nolinenumbers

\newenvironment{mybox}{\begin{tabular}{@{}l@{}}}{\end{tabular}}

\usepackage{hyperref}       

\newcommand{\eg}{\emph{e.g.}}
\newcommand{\ie}{\emph{i.e.}}

\usepackage{lmodern}

\begin{document}

\title{A $k$-mer Based Approach for SARS-CoV-2 Variant Identification}

\author{Sarwan Ali\inst{1} \and
Bikram Sahoo\inst{1} \and
Naimat Ullah\inst{2} \and
Alexander Zelikovskiy\inst{1} \and
Murray Patterson\inst{1}\thanks{To whom correspondence should be addressed.} \and
Imdadullah Khan\inst{2}$^*$
}
\authorrunning{S. Ali et al.}
%
\institute{Georgia State University, Atlanta GA, USA \\ 
\email{\{sali85,bsahoo1,alexz,mpatterson30\}@gsu.edu}
\\
\and
Lahore University of Management Sciences, Lahore, Pakistan\\
\email{\{18030048,imdad.khan\}@lums.edu.pk}}

\raggedbottom
\maketitle

\begin{abstract}

With the rapid spread of the novel coronavirus (COVID-19) across the
globe and its continuous mutation, it is of pivotal importance to
design a system to identify different known (and unknown) variants of
SARS-CoV-2.  Identifying particular variants helps to understand and
model their spread patterns, design effective mitigation strategies,
and prevent future outbreaks.  It also plays a crucial role in
studying the efficacy of known vaccines against each variant, and
modeling the likelihood of breakthrough infections.  It is well known
that the spike protein contains most of the information/variation
pertaining to coronavirus variants.

In this paper, we use spike sequences to classify different variants
of the human SARS-CoV-2.  We show that preserving order information of
the amino acids helps the underlying classifiers to achieve better
performance.  We also show that we can train our model to outperform
the baseline algorithms using only a small number of training samples
($1\%$ of the data).  Finally, we show the importance of the different
amino acids which play a key role in identifying variants and how they
coincide with those reported by the USA's Centers for Disease Control
and Prevention (CDC).

\end{abstract}

\section{Introduction}



\begin{figure}[!ht]
  \centering
  \includegraphics[scale=0.4,page=1]{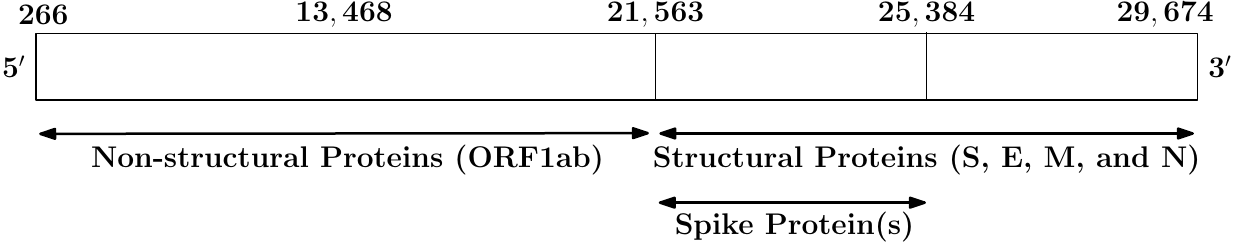}
  \caption{The SARS-CoV-2 genome is around 29--30kb encoding
    structural and non-structural proteins.  ORF1ab encodes the
    non-structural proteins and the four structural proteins: spike,
    envelope, membrane, and nucleocapsid encoded by their respective
    genes.  The spike protein has 1160 to 1542 amino acids.}
  \label{fig_spike_seq_example}
\end{figure}

The novel coronavirus, severe acute respiratory syndrome coronavirus 2
(SARS-CoV-2), that emerged in late 2019 belongs to the coronaviridae
family.  SARS-CoV-2 has an unparalleled capacity for human-to-human
transmission and became the reason for the COVID-19 pandemic.  Having
witnessed two recent pandemics caused by coronaviridae, namely
SARS-CoV in 2002 and MERS-CoV in 2012, there was an immediate research
interest in studying the zoonotic origin, transmissibility, mutation
and variants of SARS-COV-2~\cite{kuzmin2020machine,Laporte2020TheSA}.
SARS-CoV-2 has a positive-strand RNA genome of about 30kb and encodes
two categories of proteins: structural and non-structural (see
Figure~\ref{fig_spike_seq_example}).  The spike protein is one of the
substantial structural proteins of the virus, having $1160$ to $1542$
amino acids.  The spike protein's primary function is to serve as a
mechanism for the virus to enter inside the human cell by binding the
ACE2 receptor.

Detailed study of the structure of the spike glycoprotein unveils the
molecular mechanism behind host cell recognition, attachment, and
admission.  Notably, the spike glycoprotein of SARS-CoV-2 has two
subunits, $S_1$ and $S_2$, belonging to the N and C terminals,
respectively~\cite{galloway2021emergence,kuzmin2020machine}.  The
receptor binding domain (RBD) ($\approx$ $220$ amino acids) of the
$S_1$ subunit helps the virus attach to the host cell by binding the
ACE2 receptor protein, and the $S_2$ subunit helps to insert into the
cell.  SARS-CoV-2 continues to mutate over time, resulting in changes
in its amino acid sequences.  The change in the spike protein's amino
acids, specifically in the RBD, makes the virus more transmissible and
adaptable to the human immune system.  In the language of
phylogenetics and virology, the virus is creating new variants and
strains by accruing new amino acid changes in the spike protein and
its genome
\cite{galloway2021emergence,naveca2021phylogenetic,yadav2021neutralization,zhang2021emergence}.
The state-of-the-art mRNA vaccines train the host immune system to
create specific antibodies that can bind to the spike protein, which
leads to preventing the virus from entering inside the host cell.
Therefore, changing amino acids in the spike protein generates new
variants which could potentially be more contagious and more resistant
to vaccines~\cite{Krishnan2021PredictingVaccineHesitancy}.


In-depth studies of alterations in the spike protein to classify and
predict amino acid changes in SARS-CoV-2 are crucial in understanding
the immune invasion and host-to-host transmission properties of
SARS-CoV-2 and its variants.  Knowledge of mutations and variants will
help identify transmission patterns of each variant that will help
devise appropriate public health interventions to prevent rapid spread
\cite{Ahmad2016AusDM,ahmad2017spectral,Tariq2017Scalable,AHMAD2020Combinatorial}.
This will also help in vaccine design and efficacy.  A massive amount
of genomic sequences of SARS-CoV-2 are available from different parts
of the globe, with various clinical, epidemiological, and pathological
information from the GISAID
database\footnote{\url{https://www.gisaid.org/}}.  In this study, we
design sophisticated machine learning (ML) models which will leverage
the available genomic data and metadata to understand, classify and
predict the changes in amino acid in SARS-CoV-2, most notably in its
spike
protein~\cite{Krishnan2021PredictingVaccineHesitancy,Laporte2020TheSA,Lokman2020ExploringTG}.

When sequences have the same length and they are aligned, \ie, a
one-to-one correspondence between positions or indices of the
sequences is established, ML methods devised for vectors in metric
spaces can be employed for sequence analysis.  This approach treats
sequences as vectors, considering each character (\eg, amino acid or
nucleotide) as the value of the vector at a coordinate, \eg, using
one-hot encoding~\cite{kuzmin2020machine}.  In this case, the order of
positions loses its significance, however.  Since the order of indices
in sequences is a defining factor, ignoring it may result in
performance degradation of the underlying ML models.

In representation based data analysis, on the other hand, each data
object is first mapped to a vector in a fixed-dimensional vector
space, taking into account important structure in the data (such as
order).  Vector space ML algorithms are then used on the vector
representations of sequences.  This approach has been highly
successful in the analysis of data from various domains such as
graphs~\cite{hassan2020estimating,Hassan2021Computing}, nodes in
graphs~\cite{ali2021predicting}, and electricity
consumption~\cite{ali2019short,Ali2020ShortTerm}.
This approach yields significant success in sequence analysis, since
the feature representation takes into account the sequential nature of
the data, such as
texts~\cite{Shakeel2020LanguageIndependent,Shakeel2020Multi,Shakeel2019MultiBilingual},
electroencephalography and electromyography
sequences~\cite{atzori2014electromyography,ullah2020effect},
Networks~\cite{Ali2019Detecting1}, and biological
sequences~\cite{leslie2002mismatch,farhan2017efficient,Kuksa_SequenceKernel,ali2021effective,ali2021simpler,ali2021spike2vec}.
For biological sequences (DNA and protein), a feature vector based on
counts of all length $k$ substrings (called $k$-mers) occurring
exactly or inexactly up to $m$ mismatches (mimicking biological
mutations) is proposed in~\cite{leslie2002mismatch}.  The kernel value
between two sequences --- the dot product between the corresponding
feature vectors --- serves as a pairwise proximity measure and is the
basis of kernel based ML.  We provide the technical definition of
feature maps and the computational aspects of kernel computation in
Section~\ref{sec_proposed_approach}.  In this paper, our contributions
are the following:
\begin{enumerate}
\item We propose a method based on $k$-mers (for feature vector
  generation) and kernel approximation (to compute pairwise similarity
  between spike sequences) that classify the variants with very high
  accuracy.
\item We show that spike sequences alone can be used to efficiently
  classify different COVID-19 variants.
\item We show that the proposed method outperforms the baseline in
  terms of accuracy, precision, recall, F1-measure, and ROC-AUC.
\item We show that with only $1\%$ of the data for training, we can
  achieve high prediction accuracy.
\end{enumerate}

The rest of the paper is organised as follows:
Section~\ref{related_work} contains the previous work related to our
problem.  Section~\ref{sec_proposed_approach} contains the proposed
approach of this paper in detail, along with the description of the
baseline model.  Section~\ref{sec_data_set_detail} contains the
information related to different datasets.  We present our results in
Section~\ref{sec_results_and_discussion}. Finally, we conclude the
paper in Section~\ref{sec_conclusion}.








\section{Related Work }\label{related_work}




Phylogeny based inference of disease transmission~\cite{Dhar2020TNet}
and sequence homology (shared ancestry) detection between a pair of
proteins are important tasks in bioinformatics and biotechnology.
Sequence classification is a widely studied problem in both of these
domains~\cite{Krishnan2021PredictingVaccineHesitancy}.  Most sequence
classification methods for viruses require the alignment of the input
sequence to fixed length predefined vectors, which enables the machine
learning algorithm to compare homologous feature
vectors~\cite{farhan2017efficient}.  Pairwise local and global
alignment similarity scores between sequences were used traditionally
for sequence analysis.  Alignment-based methods are computationally
expensive, especially for long sequences, while heuristic methods
require a number of ad-hoc settings such as the penalty of gaps and
substitutions, and alignment methods may not perform well on highly
divergent regions of the genome.  To address these limitations,
various alignment-free classification methods have been proposed
\cite{farhan2017efficient}.
The use of $k$-mer (substrings of length $k$) frequencies for
phylogenetic applications started
with~\cite{Blaisdell1986AMeasureOfSimilarity}, which reported success
in constructing accurate phylogenetic trees from several coding and
non-coding nDNA sequences.  Typically, $k$-mer frequency vectors are
paired together with a distance function to measure the quantitative
similarity score (kernel matrix) between any pair of
sequences~\cite{farhan2017efficient}.  However, the basic bottleneck
for these techniques is the quadratic (in the lengths of the
sequences) runtime of kernel evaluation.



With the global outbreak of Covid-19 in 2020, different mutations of
its variants were discovered by the genomics community.  Massive
testing and large-scale sequencing produced a huge amount of data,
creating ample opportunity for the bioinformatics community.
Researchers started exploring the evolution of
SARS-CoV-2~\cite{Ewen2021TargetedSelf,melnyk2020clustering} to vaccine
landscapes~\cite{su2021learning} and long-term effects of covid to
patients~\cite{tankisi2020critical}.  In~\cite{Laporte2020TheSA}, the
authors indicate how the coronavirus spike protein is fine-tuned
towards the temperature and protease conditions of the airways, to
enhance virus transmission and pathology.
After the global spread of the coronavirus, researchers started
exploring ways to identify new variants and measuring vaccine
effectiveness.  In~\cite{Leila2021Genotype}, the authors study genome
structures of SARS-CoV-2 and its previous versions,
while~\cite{Lokman2020ExploringTG} explores the genomics and proteomic
variants of the SARS-CoV-2 spike glycoprotein.





\section{Proposed Approach}\label{sec_proposed_approach}


Given a set of spike sequences, our goal is to find the similarity
score between each pair of sequences (kernel matrix).
The resultant kernel matrix is given as input to the kernel PCA method
for dimensionality reduction.  The reduced dimensional principal
components-based feature vector representation is given as input to
the classical machine learning models.  We discuss each step in detail
below.

\subsection{$k$-mers Computation}

\begin{figure}[!ht]
  \centering
  \includegraphics[scale = 0.3] {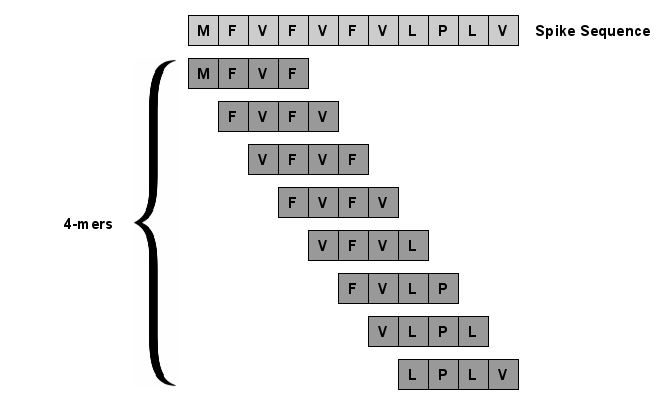}
  \caption{Example of $k$-mers (where k = 4) in a spike sequence
    "MFVFVFVLPLV".}
  \label{fig_k_mer_demo}
\end{figure}

For mapping protein sequences to fixed-length vectors, it is important
to preserve order information of the amino acids within a sequence.
To achieve this, we use substrings (called mers) of length $k$.  For
each spike sequence, the total number of $k$-mers are "$N - k + 1$",
where $N$ is the total number of amino acids in the spike sequence
($1274$), and $k$ is a user-defined parameter for the size of each
mer.  An example of $k$-mers (where k = $4$) is given in Figure
\ref{fig_k_mer_demo}.

However, since we do not know the total unique $k$-mers in all spike
sequences, we need to consider all possible pairs of $k$-mers to
design a general purpose feature vector representation for spike
sequences in a given dataset.  In this way, given an alphabet $\Sigma$
(a finite set of symbols), we know that a spike sequence $X \in
\Sigma$ ($X$ contains a list of ``ordered'' amino acids from
$\Sigma$).  Similarly, we can extract sub-strings (mers) from $X$ of
length $k$, which we called $k$-mers.

Given $X$, $k$, and $\Sigma$, we have to design a frequency feature
vector $\Phi_k (X)$ of length $\vert \Sigma \vert^k$, which will
contain the exact number of occurrences of each possible $k$-mer in
$X$.  The distance between two sequences $X$ and $Y$ is then simply
the hamming distance $d_H$ (count the number of mismatched values).
After computing the feature vector, a kernel function is defined that
measures the pairwise similarity between pairs of feature vectors
(usually using dot product).  The problem we consider so far is the
huge dimensionality of the feature vector $\vert \Sigma \vert^k$ that
can make kernel computation very expensive.  Therefore, in the
so-called {\em kernel trick}, kernel values are directly evaluated
instead of comparing indices.



The algorithm proposed in \cite{farhan2017efficient} takes the feature
vectors (containing a count of each $k$-mers) as input and returns a
real-valued similarity score between each pair of vectors.  Given two
feature vectors $A$ and $B$, the kernel value for these vectors is
simply the dot product of $A$ and $B$.  For example, given a $k$-mer,
if the frequency of that $k$-mer in $A$ is $2$ and $B$ is $3$, its
contribution towards the kernel value of $A$ and $B$ is simply $2
\cdot 3$.  The process of kernel value computation is repeated for
each pair of sequences and hence we get a (symmetric) matrix (kernel
matrix) containing a similarity score between each pair of sequences. Note that $k$ is a user-defined parameter --- in our experiments we
use $k = 9$.

\begin{theorem}\label{thm_kernel}~\cite{farhan2017efficient}
  The runtime of the kernel computation is bounded above by $O(k^2 \ n
  \ log \ n)$, where $k$ is the length of $k$-mers and $n$ is the
  length of sequences.
\end{theorem}

\subsection{Kernel PCA}

Due to a high-dimensional kernel matrix, we use Kernel PCA (K-PCA)
\cite{hoffmann2007kernel} to select a subset of principal components.
These extracted principal components corresponding to each spike
sequence act as the feature vector representations for the spike
sequences (we selected 50 principal components for our experiments).

\subsection{Machine Learning Classifiers}

Various ML algorithms have been utilized for the
classification task.  K-PCA output, which is $50$ components fed to
different classifiers for prediction purposes.  We use Support Vector
Machine (SVM), Naive Bayes (NB), Multi-Layer Perceptron (MLP),
K-Nearest Neighbour (KNN) (with $K = 5$), Random Forest (RF), Logistic
Regression (LR), and Decision Tree (DT) classifiers.  
All experiments are done on a Core i5 system with Windows 10 OS and 32
GB RAM.  Implementation of our algorithm is done in Python. Our code
and pre-processed datasets are available
online \footnote{\url{https://github.com/sarwanpasha/covid_variant_classification}}.
The evaluation metrics that we are using are average accuracy, precision, recall, weighted and macro F1, and ROC area under the curve (AUC).

\subsection{Baseline Model}

We consider the approach of~\cite{kuzmin2020machine} as a baseline
model.  The authors of~\cite{kuzmin2020machine} convert spike
sequences into one-hot encoding vectors that are used to classify the
viral hosts. We have the 21 amino acids
"\textit{ACDEFGHIKLMNPQRSTVWXY}" (unique alphabets forming $\Sigma$).
The length of each spike sequence is $1273$ (with $*$ at the
$1274^{th}$ location).
After converting sequences into one-hot encoding vectors we will get a
$26,733$ dimensional vector ($21 \times 1273 = 26,733$).  Principal
Component Analysis (PCA) on these vectors is applied to reduce
dimensionality for the underlying classifiers.
For reference, we use the name ``One-Hot'' for this baseline approach
in the rest of the paper. For PCA, we select $100$ principal
components.


\section{Dataset Description and Preprocessing}
\label{sec_data_set_detail}

We sampled three subsets of spike sequences from the largest known
database of human SARS-CoV-2,
GISAID \footnote{\url{https://www.gisaid.org/}}.  We refer to those
$3$ subsets as GISAID 1, GISAID 2, and GISAID 3, having $7000$,
$7000$, and $3029$ (aligned) spike sequences, respectively, each of
length 1274 from $5$ variants.  For GISAID 1 and GISAID 2 datasets, we
preserve the proportion of each variant as given in the original
GISAID database. For the GISAID 3 dataset, we use a comparatively
different proportion of variants to study the behavior of our
algorithm (see Table~\ref{tbl_variant_information}).





\begin{table}[H]
  \centering
  \begin{tabular}{lllc | l}
    \hline
    \begin{mybox}
      Pango\\Lineage\end{mybox} & Region & Labels & \begin{mybox}
	Num mutations\\S-gene/Genome  \end{mybox} & \begin{mybox}\hskip.3in Num sequences in\\ GISAID 1 \hskip.01in GISAID 2 \hskip.01in GISAID 3 \end{mybox} \\
      \hline	\hline	
      B.1.1.7 & UK~\cite{galloway2021emergence} &  Alpha & 8/17 & \hskip.1in 5979	\hskip.3in 5979\hskip.4in 2055\\
      B.1.351  & South Africa~\cite{galloway2021emergence}  &  Beta & 9/21& \hskip.1in 124
      \hskip.4in 124\hskip.4in 133\\
      P.1  &  Brazil~\cite{naveca2021phylogenetic} &  Gamma &  10/21 & \hskip.1in 202
      \hskip.4in 202\hskip.4in 625\\
      B.1.617.2  & India~\cite{yadav2021neutralization}  &  Delta &  8/17  & \hskip.1in 596
      \hskip.4in 596\hskip.4in 44\\
      B.1.427   & California~\cite{zhang2021emergence}  & Epsilon  &  3/5 & \hskip.1in 99
      \hskip.5in 99\hskip.4in 182\\
      

        
       
      \hline
  \end{tabular}
  \caption{Variants information and distribution in the three
    datasets. The S/Gen. column represents number of mutations on the
    S gene / entire genome.}
  \label{tbl_variant_information}
\end{table}
\vskip-.3in

\begin{figure}[!ht]
  \centering
  \includegraphics[scale = 0.55,page = 1] {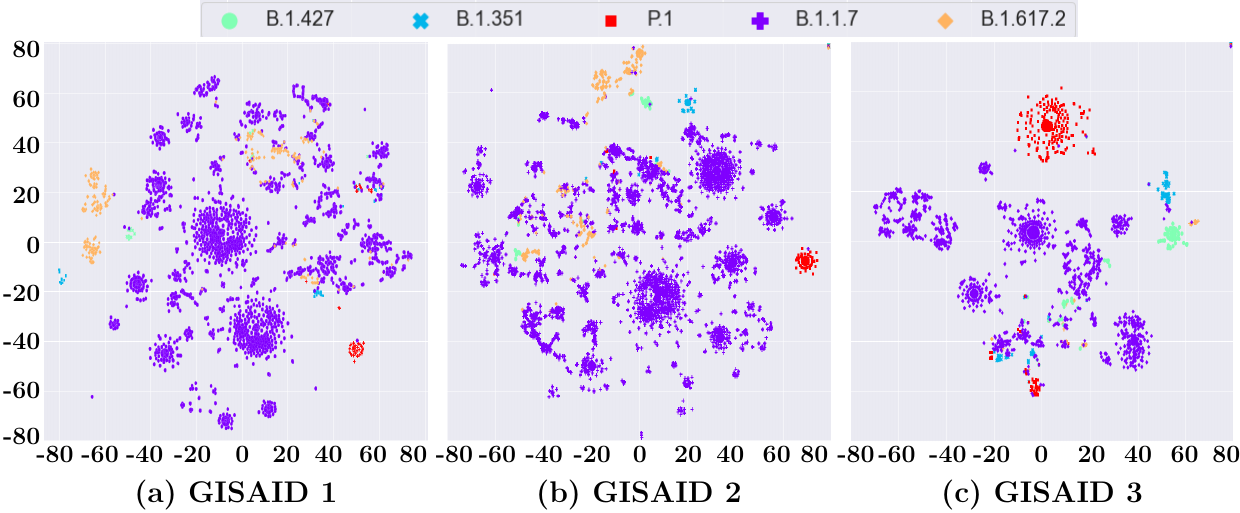}
  \caption{t-SNE embeddings of spike sequences.}
  \label{fig_tsn_embedding}
\end{figure}

\noindent To visualize the local structure of spike sequences, we use
t-distributed stochastic neighbor embedding (t-SNE)
\cite{van2008visualizing} that maps input sequences to 2D real
vectors.  The t-SNE helps to visualize (hidden) clusters in the data.
The visualization results are shown in Figure~\ref{fig_tsn_embedding},
revealing that variants are not well separated, making variant
classification a challenging task. It is clear from
Figure~\ref{fig_tsn_embedding} that the dominant Alpha variant is not
in a single cluster, and smaller variants are scattered around (\eg,
the least frequent variant, B.1.427, appears in most clusters).

\section{Experimental Evaluation}
\label{sec_results_and_discussion}

In this section, we first report the performance of different
classifiers using multiple performance metrics.  Then we analyze the
importance of the positions of each amino acid in the spike sequence
using information gain.  Results for the GISAID 1, 2 and 3 datasets
are given in
Tables~\ref{tbl_avg_classification_results_second_dataset}--\ref{tbl_avg_classification_results_third_dataset}.
We present results for each classifier separately for the baseline
method and compare it with our proposed method.  We can observe that
for most of the classifiers, our proposed method is better than the
baseline.  For example, in the case of SVM classifier, the One-Hot
method got $0.962$ F1-Macro score for the GISAID 1 dataset while our
proposed model got $0.973$, which is a significant improvement
considering that all values are on the higher side.  Similar behavior
is observed for other classifiers also.  For all of these results, we
use $1\%$ data for training and $99\%$ for testing purposes.  Since we
are getting such high accuracies, we can conclude that with a minimum
amount of available data, we can train a classifier that can classify
different variants very efficiently.  Also, we can observe that the
SVM classifier is consistently performing best for all the datasets.
Note that results in
Tables~\ref{tbl_avg_classification_results_second_dataset}--\ref{tbl_avg_classification_results_third_dataset}
are averaged over all variants.


\begin{table}[!ht]
  \centering
  \begin{tabular}{cp{0.8cm}cccccc}
    \hline
    Approach & ML Algo. & Acc. & Prec. & Recall & F1 (weighted) & F1 (Macro) & ROC-AUC \\	
    \hline	\hline		
    \multirow{7}{*}{One-Hot \cite{kuzmin2020machine}} 
    & SVM & 0.990 & 0.990 & 0.990 & 0.990 & 0.962 & 0.973 \\
    & NB & 0.957 & 0.964 & 0.951 & 0.952 & 0.803 & 0.881 \\
    & MLP & 0.972 & 0.971 & 0.975 & 0.974 & 0.881 & 0.923 \\
    & KNN & 0.978 & 0.964 & 0.977 & 0.965 & 0.881 & 0.900 \\
    & RF & 0.964 & 0.962 & 0.961 & 0.963 & 0.867 & 0.878 \\
    & LR & 0.985 & 0.981 & 0.983 & 0.984 & 0.935 & 0.950 \\
    & DT & 0.941 & 0.945 & 0.947 & 0.944 & 0.793 & 0.886\\
    \hline
    \multirow{7}{*}{Kernel Approx.} 
    & SVM & \textbf{0.994} & \textbf{0.994} & \textbf{0.995} & \textbf{0.995} & \textbf{0.973} & \textbf{0.988} \\
    & NB & 0.987 & 0.985 & 0.985 & 0.986 & 0.901 & 0.912 \\
    & MLP & 0.975 & 0.977 & 0.976 & 0.978 & 0.921 & 0.935 \\
    & KNN & 0.979 & 0.967 & 0.979 & 0.967 & 0.887 & 0.904 \\
    & RF & 0.981 & 0.987 & 0.988 & 0.980 & 0.944 & 0.945 \\
    & LR & 0.992 & 0.990 & 0.993 & 0.992 & 0.991 & 0.990 \\
    & DT & 0.985 & 0.981 & 0.985 & 0.987 & 0.898 & 0.944\\ 
    \hline
  \end{tabular}
  \caption{Variants Classification Results (1\% training set and 99\%
    testing set) for the GISAID 1 Dataset. Best values are shown in
    bold.}
  \label{tbl_avg_classification_results_second_dataset}
\end{table}

\begin{table}[!ht]
  \centering
  \begin{tabular}{cp{0.8cm}cccccc}
    \hline
    Approach & ML Algo. & Acc. & Prec. & Recall & F1 (weighted) & F1 (Macro) & ROC-AUC \\	
    \hline	\hline		
    \multirow{7}{*}{One-Hot \cite{kuzmin2020machine}} 
    & SVM & 0.994 & 0.994 & 0.993 & 0.992 & 0.975 & 0.983 \\
    & NB & 0.912 & 0.936 & 0.912 & 0.920 & 0.794 & 0.913 \\
    & MLP & 0.970 & 0.970 & 0.970 & 0.969 & 0.880 & 0.921 \\
    & KNN & 0.960 & 0.960 & 0.960 & 0.958 & 0.841 & 0.863 \\
    & RF & 0.966 & 0.967 & 0.966 & 0.964 & 0.888 & 0.885 \\
    & LR & 0.993 & 0.993 & 0.993 & 0.993 & 0.968 & 0.973 \\
    & DT & 0.956 & 0.957 & 0.956 & 0.956 & 0.848 & 0.913 \\
    \hline
    \multirow{7}{*}{Kernel Approx.} 
    & SVM & \textbf{0.998} & \textbf{0.997} & \textbf{0.997} & \textbf{0.998} & \textbf{0.998} & \textbf{0.997} \\
    & NB & 0.985 & 0.988 & 0.985 & 0.984 & 0.946 & 0.967 \\
    & MLP & 0.973 & 0.971 & 0.972 & 0.970 & 0.889 & 0.925 \\
    & KNN & 0.965 & 0.962 & 0.963 & 0.967 & 0.845 & 0.867 \\
    & RF & 0.990 & 0.992 & 0.991 & 0.996 & 0.978 & 0.977 \\
    & LR & 0.997 & 0.994 & 0.996 & 0.997 & 0.991 & 0.993 \\
    & DT & 0.991 & 0.990 & 0.994 & 0.996 & 0.952 & 0.963 \\
    \hline
  \end{tabular}
  \caption{Variants Classification Results (1\% training set and 99\%
    testing set) for the GISAID 2 Dataset. Best values are shown in
    bold.}
  \label{tbl_avg_classification_results_first_dataset}
\end{table}



\begin{table}[!ht]
  \centering
  \begin{tabular}{cp{0.8cm}cccccc}
    \hline
    Approach & ML Algo. & Acc. & Prec. & Recall & F1 (weighted) & F1 (Macro) & ROC-AUC \\	
    \hline	\hline		
    \multirow{7}{*}{One-Hot \cite{kuzmin2020machine}}  & SVM & 0.988 & 0.986 & 0.987 & 0.982 & 0.924 & 0.961 \\
    & NB & 0.764 & 0.782 & 0.761 & 0.754 & 0.583 & 0.747 \\
    & MLP & 0.947 & 0.941 & 0.944 & 0.942 & 0.813 & 0.898 \\
    & KNN & 0.920 & 0.901 & 0.924 & 0.901 & 0.632 & 0.773 \\
    & RF & 0.928 & 0.935 & 0.922 & 0.913 & 0.741 & 0.804 \\
    & LR & 0.982 & 0.981 & 0.983 & 0.984 & 0.862 & 0.921 \\
    & DT & 0.891 & 0.891 & 0.890 & 0.895 & 0.679 & 0.807\\ 
    \hline
    \multirow{7}{*}{Kernel Approx.} 
    & SVM & \textbf{0.991} & \textbf{0.993} & \textbf{0.995} & \textbf{0.991} & \textbf{0.989} & \textbf{0.997} \\
    & NB & 0.864 & 0.922 & 0.861 & 0.884 & 0.783 & 0.887 \\
    & MLP & 0.926 & 0.922 & 0.921 & 0.923 & 0.805 & 0.909 \\
    & KNN & 0.947 & 0.921 & 0.942 & 0.934 & 0.701 & 0.826 \\
    & RF & 0.975 & 0.971 & 0.971 & 0.972 & 0.904 & 0.918 \\
    & LR & 0.991 & 0.990 & 0.994 & 0.990 & 0.983 & 0.992 \\
    & DT & 0.960 & 0.969 & 0.964 & 0.967 & 0.812 & 0.891\\ 
    \hline
  \end{tabular}
  \caption{Variants Classification Results (1\% training set and 99\%
    testing set) for the GISAID 3 Dataset. Best values are shown in
    bold.}
  \label{tbl_avg_classification_results_third_dataset}
\end{table}


We also show the variant-wise performance of our best classifier
(SVM).  Table~\ref{tbl_svm_heatmap} contains the resulting confusion
matrices using the Kernel Approx. and One-Hot approaches for GISAID
1. Clearly, the kernel-based approach performs better than the one-hot
approach for most of the variants.

\begin{table}[h!]
  \footnotesize
  \subfloat{
    \begin{tabular}{@{}l|cccccccc@{}}
      \toprule
      Variant & {\bf Alpha} & {\bf Beta} & {\bf Delta} & {\bf Gamma} & {\bf Epsi.}  \\ \midrule
      {\bf Alpha} & 5373 & 3 & 7 & 0 & 5  \\ 
      {\bf Beta} & 6 & 110 & 0 & 0 & 0   \\
      {\bf Delta} & 6 & 0 & 523 & 0 & 0  \\ 
      {\bf Gamma} & 0 & 0 & 0 & 176 & 0  \\ 
      {\bf Epsilon} & 2 & 0 & 0 & 0 & 89 \\ 
      \bottomrule 
    \end{tabular}
  }
  \subfloat{
    \footnotesize
    \begin{tabular}{@{}|cccccccc@{}}
      \toprule
          {\bf Alpha} & {\bf Beta} & {\bf Delta} & {\bf Gamma} & {\bf Epsi.}  \\ \midrule
          5371 & 9 & 5 & 0 & 3  \\ 
          13 & 103 & 0 & 0 & 0   \\ 
          8 & 0 & 521 & 0 & 0   \\ 
          0 & 0 & 0 & 176 & 0  \\ 
          7 &  0 & 3 & 0 & 81 \\ 
          \bottomrule 
    \end{tabular}
  }
  \caption{Confusion matrices for SVM classifiers using Kernel
    Approx. approach (left) and using One-Hot approach (right) for the
    GISAID 1 dataset.}
  \label{tbl_svm_heatmap}
\end{table}

\vskip-.2in

\subsection{Importance of Amino Acid Positions}

\begin{figure}[t]
  \centering
  \centering
  \includegraphics{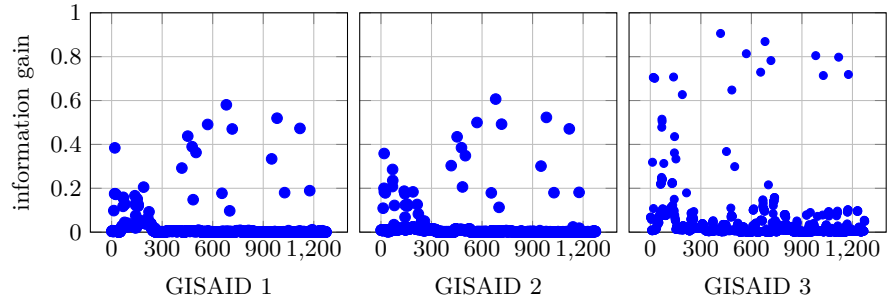}
  \caption{Information gain of each amino acid position with respect
    to variants. The $x$-axis corresponds to amino acid positions in
    the spike sequences.}
  \label{fig_IG_dataset_1_2_3}
\end{figure}

To evaluate each individual position's importance, we measure the Information Gain (IG) for each position with respect to the variant defined as
$IG(Class,position) = H(Class) - H(Class | position)$, where $H=
\sum_{ i \in Class} -p_i \log p_i$ is the entropy, and $p_i$ is the
probability of the class $i$.  Figure~\ref{fig_IG_dataset_1_2_3}
depicts how informative a position is to determine variants (higher
value is better).  
We observe that positions such as 452, 570 and 681 are more informative across all datasets.  The USA's CDC
also declared mutations at these positions from one variant to the
other, which validates our feature selection algorithm.  For instance,
R452L is present in B.1.427(Epsilon) and B.1.617 (Kappa, Delta)
lineages and sub-lineages.  
The combination of K417N, E484K, and N501Y substitutions are present in B.1.351 (Beta).  
Similarly, K417T, E484K,
and N501Y substitutions are present in
P.1(Gamma)\footnote{\url{https://www.cdc.gov/coronavirus/2019-ncov/variants/variant-info.html}}
(they can be seen having higher IG in
Figure~\ref{fig_IG_dataset_1_2_3}).



\section{Conclusion and Future Directions}
\label{sec_conclusion}


We propose an approach to efficiently classify SARS-CoV-2 variants
using spike sequences.  Results show that the $k$-mer based sequence
representation outperforms the typical one-hot encoding approach since
it preserves order information of amino acids.  We showed the
importance of specific amino acids and demonstrate that it agrees with
the CDC definitions of variants.  In the future, we will work towards
detecting new (unknown) variants based on whole genome sequences.
Another exciting future work is considering other attributes like
countries, cities, and dates to design richer feature vector
representations for spike sequences.

\bibliographystyle{splncs04}
\bibliography{isbra}
\end{document}